\documentclass[a4paper,11pt]{article}

\usepackage{jheppub} 

\usepackage[T1]{fontenc} 

     \usepackage{bm}
\usepackage{color}

\newcommand{\Ds}{\displaystyle}                           
 \renewcommand{\thesection}{\arabic{section}}

\renewcommand{\thetable}{\arabic{table}}
\renewcommand{\thefigure}{\arabic{figure}}

\renewcommand{\tablename}{Table}
\renewcommand{\figurename}{Figure}

\newcommand{\be}{\begin{equation}}\newcommand{\ee}{\end{equation}}%
\newcommand{\bd}{\begin{displaymath}}\newcommand{\ed}{\end{displaymath}}
\newcommand{\bit}{\begin{itemize}}                        
 \newcommand{\eit}{\end{itemize}}                         
\newcommand{\ben}{\begin{enumerate}}                      
 \newcommand{\een}{\end{enumerate}}                       
\newcommand{\baa}{\begin{array}{lll}}                     
 \newcommand{\eaa}{\end{array}}                           
\newcommand{\ba}{\begin{eqnarray}}                        
 \newcommand{\ea}{\end{eqnarray}}                         
\newcommand{\gev}[1]{\relax\ifmmode{\text{GeV}^{#1}}      
                     \else{GeV$^{#1}${ }}\fi}             
\newcommand{\Gev}{\relax\ifmmode{\text{GeV}}              
                     \else{GeV{ }}\fi}                    
\newcommand{\Mev}{\relax\ifmmode{\text{MeV}}              
                     \else{MeV{ }}\fi}                    
\def\MSbar{\relax\ifmmode\overline                        
            {\rm MS}\else{$\overline{\rm MS}${ }}\fi}     
\def\as{\relax\ifmmode \alpha_s\else{$ \alpha_s${ }}\fi}  
\def\abar{\relax\ifmmode{\bar{a}}\else{$\bar{a}${ }}\fi}  
  \def\ie{\hbox{\it i.e.}{ }} 
   \def\eg{\hbox{\it e.g.}{ }}  

\def\1{\hbox{{1}\kern-.25em\hbox{l}}}

\title{\boldmath The   $\{\beta\}$-expansion formalism in perturbative QCD and its extension
}
 \author[a,b]{A.~L. Kataev}
  \author[c]{S.~V.~Mikhailov}

\affiliation[a]{Institute for Nuclear Research of the Academy
of Sciences of Russia, 117312, Moscow, Russia}
   \affiliation[b]{Moscow Institute of Physics and Technology, 141700, Dolgoprudny, Moscow Region, Russia}
   \affiliation[c]{Bogoliubov Laboratory of Theoretical Physics, JINR,
                   141980,  Dubna, Russia}

\emailAdd{kataev@ms2.inr.ac.ru}
\emailAdd{mikhs@theor.jinr.ru}
\abstract{
We discuss  the $\{ \beta \}$-expansion for renormalization group invariant
quantities tracing this expansion to the different contractions of the corresponding incomplete
BPHZ $R$-operation.
 All of the coupling renormalizations, which follow from these  contractions, should be
  taken into account for the $\{ \beta \}$-expansion.
 We illustrate this  feature
considering the
nonsinglet Adler function $D^\text{NS}$ in
 the third order of perturbation.
We propose a generalization of the $\{ \beta \}$-expansion for the renormalization group
covariant  quantities -- the $\{ \beta,\gamma \}$-expansion.
}

\begin{document}
 \maketitle

\section{Introduction}
In the last few years, interest in the construction of the
high-order generalizations of the  Brodsky-Lepage-Mackenzie (BLM) approach
\cite{Brodsky:1982gc} for fixing the scales in
the perturbative QCD expansions of the  renormalization-group invariant (RGI)
quantities in powers of the coupling constant
 $\alpha_s(\mu^2)$ was renewed
(see, e.g., \cite{MS04,Wu:2013ei,Gehrmann:2014uva, Kataev:2014jba,
Kataev:2014zwa}).
 A special feature of the new class of the BLM based procedures
is the  absorption into new  scales  of  those  terms that are  proportional
to the coefficients  of the  QCD $\beta$-function fill factors.
 These new perturbative expansions are based on  the expansion of
the terms of the massless perturbative QCD series for the   RGI-invariant
quantities in the powers and  the products  of powers of the coefficients
of the QCD $\beta$-function.
It is   called the   $\{\beta\}$-expansion formalism
(see, e.g., \cite{Kataev:2014jba}) and   was  first proposed  in  \cite{MS04}
as the double series representation for the quantity in the parameters
$\alpha^n_s$ and $\beta^l_i$ -- the powers of the $\beta$-function
 coefficients (``matrix representation'').
Then
this $\{\beta\}$-expansion was intensively
 studied
in 
 \cite{Kataev:2010du, Kataev:2014jba,Kataev:2014zha, Kataev:2014zwa, Bakulev:2010gm}
 in the case of QCD supplemented
with multiplets  of gluions, which are the elements of general
SUSY extension
of  QCD theory.
In 
\cite{Brodsky:2011ta, Mojaza:2012mf, Brodsky:2013vpa},
the variants of the $\{\beta\}$-expansion were used in QCD without additional degrees
of freedom, while in \cite{Ma:2015dxa} the variants
of the $\{\beta\}$-expansion procedure were applied in both the cases, namely
in QCD with and without additional gluino degrees of freedom.
Note that in general the $\{\beta\}$-expansion is  formulated
for the   RGI quantities, initially evaluated within the class of
minimal subtraction (MS)-schemes  (see, e.g.,  \cite{MS04,Mojaza:2012mf}).

At the next step of elaboration of the modern generalizations of the BLM approach  all the terms,
proportional to the coefficients of the QCD $\beta$-function,
are   transferred into the  scales $\mu^2$ of the powers
of $\alpha_s(\mu^2)$    (or   $a_s(\mu^2)=\alpha_s(\mu^2)/(4\pi)$)
in the corresponding series
(see,  e.g., \cite{MS04,Brodsky:2011ta,Mojaza:2012mf,Brodsky:2013vpa,Ma:2015dxa}).
However, there are still several  points of view
how to use the BLM approach that is  generalized in this way
and how to fix the concrete content of the $\{\beta\}$-dependent elements in
the perturbative coefficients in the framework of the MS-like renormalization scheme.
They appear during the   development and   study of  the applications
of  the  BLM  generalizations in practice.
They are pushed forward  mainly by two groups: the authors of the present work
and the authors of the ``Principle of Maximum Conformality'' (PMC) also   based
on  the  $\{\beta\}$-expansion \cite{Brodsky:2011ta,Mojaza:2012mf,Brodsky:2013vpa}.
 The first  theoretical disagreement finally leads
to different values of the $\beta $-dependent elements
of the expansion coefficients of the RGI quantities
\footnote{
 In the recent paper  the general
form of the  $\{\beta\}$-expended expressions for these  coefficients
\textit{  is the same}  as  the one,
introduced in \cite{MS04} and used
in \cite{Kataev:2010du,Kataev:2014jba}, but the
concrete coefficients remain different.}
to different results for new BLM-type scales,
and to different points of view on the  scheme-dependence of the
$\{\beta\}$-independent terms in the certain PMC-type series.
The second disagreement is related to different interpretations  of
the $\{\beta\}$-expansion for two  related representations of the
$e^+e^-$-annihilation Adler function $D^\text{EM}\left(a_s\right)$ which,
contrary to the considerations presented  in \cite{Wu:2013ei, Brodsky:2013vpa, Ma:2015dxa},
 should lead to the identical results,
in full agreement with the
basis of the incomplete   BPHZ R-operation described in detail in \cite{BSHbook}.

The first part of this work is devoted to the  proof of the latter statement.
In Sec.\ref{Sec:2}
we discuss the definitions and properties  of
the photon vacuum polarization function $\Pi^\text{EM}(a_s)$ and  of  its anomalous
dimension $\gamma_{ph}^{EM}(a_s)$  following the detailed considerations in
\cite{Chetyrkin:1980sa,Chetyrkin:1980pr}.
In Sec. \ref{Sec:3} we recall to readers the details
of the calculations of $D^\text{EM}$.
The aim is to demonstrate how the definition of the $\{\beta\}$-expansion
 proposed in \cite{MS04} can be realized  for the case of the
$O(a_s^4)$   representation of the $D^\text{EM}$-function explicitly presented in  \cite{Baikov:2012zm}
 in terms of the photon anomalous dimension and the
polarization function $\Pi^\text{EM}(a_s)$,
which was used in the consideration of \cite{Brodsky:2013vpa}.
We shall  clarify the statement  already made in \cite{Kataev:2014jba}
that the careful application of the
$\{\beta\}$-expansion in this case  leads to the same results for
the $\{\beta\}$-expansion obtained in \cite{MS04}
without involving into consideration this
presentation for the $D^\text{EM}$-function
with the photon anomalous dimension as well.
This clarification should be compared with the  non-completed
analysis presented in \cite{Wu:2013ei,Brodsky:2013vpa,Ma:2015dxa}.

In  Section \ref{Sec: R'-operation},
we demonstrate that the $\{\beta\}$-expansion  of the
RGI-invariant quantities (in the MS-like schemes)  can be
traced to the $R'$-operation, i.e. the incomplete BPHZ R-operation.
We recall, following
the  studies in \cite{Chetyrkin:1980sa,Chetyrkin:1980pr},
that the subtraction of all QCD  subdivergences from the bare  photon vacuum polarization function,
 which enter into  the definition of the $D^\text{EM}$-function,
is equivalent to the renormalization of the bare QCD coupling constant  $a_{sB}$ in this expression.
This allows us to show  in a more formal way why  the application of
the $\{\beta\}$-expansion approach to the different presentations
of the Adler function through  $\gamma_{ph}^{EM}(a_s)$ and $\Pi^\text{EM}(a_s)$ and
through the   K\"all\'en-Lehmann  dispersion representation
give  identical results.

In Section \ref{Sec:beta-gamma}, we generalize the $\{\beta\}$-expansion
to the case  of the RG-covariant
quantities, which  have their own anomalous dimensions.
In   this case, the structure of the expansion of the coefficient will
differ from the structure of the $\{\beta\}$-expansion of the RG-invariant quantity
-- it will  contain new terms,
which are proportional to the coefficients of the corresponding anomalous dimension.
The combined consideration of these terms forms the basis for the new $\{\beta, \gamma\}$-expansion.
We emphasize the necessity of further development of
the optimization of the related  series in the spirit of the generalized  BLM approach.

\section{The Adler function  and its $\{\beta\}$--expansion}
\label{Sec:2}
Let us start  first with the expression for the
vacuum polarization tensor, which is  related to
the polarization function $\Pi^\text{EM}$ as
\begin{equation}
\label{Pimunu}
\Pi_{\mu\nu}^\text{EM}(q,a_s)=i \int e^{iqx}  \langle 0|J^\text{EM}_{\mu}(x)J^\text{EM}_{\nu}(0)|0\rangle d^dx
= \bigg(q_{\mu}q_{\nu}-g_{\mu\nu}q^2\bigg)\Pi^\text{EM}(L,a_s)\,.
\end{equation}
Here   $J^\text{EM}_{\mu}(x)=\sum_i q_i\cdot \overline{\psi}_{i}\gamma_{\mu}\psi_{i}$, $q_i$ are
the electromagnetic charges of quarks, $-q^2=Q^2$, $L=\ln(\mu^2/Q^2)$.
Due to the vector current conservation  the r.h.s.  of
Eq.(\ref{Pimunu})  is transverse.
The vacuum polarization function $\Pi^\text{EM}(L,a_s)$ contains
the non-singlet (NS) and
singlet (SI) contributions
\begin{equation}
\label{NS-SI}
\Pi^\text{EM}(L,a_s)=\bigg(\sum_i q_i^2\bigg)\Pi^\text{NS}(L,a_s)+
\bigg(\sum_i q_i\bigg)^2\Pi^\text{SI}(L,a_s)\,.
\end{equation}
The latter ones appear   for the first time at the four-loop level
\cite{Gorishnii:1990vf,SurgSam:1990tg,Chetyrkin:1996ez}.
The Adler function $D^\text{EM}$ is a widely used characteristic of
the $e^+e^-$-annihilation to a hadrons process,
namely
\begin{equation}
\label{D-function}
D^\text{EM}(L,a_s)=-12\pi^2 Q^2\frac{d}{dQ^2}\Pi^\text{EM}(L,a_s)=Q^2\int_{4m_\pi^2}^{\infty}
\frac{R(s)}{(s+Q^2)^2} ds\,,
\end{equation}
which was introduced in \cite{Adler:1974gd} in the Euclidean region.
In the RHS  of this
K\"all\'en-Lehmann representation the spectral function
$R(s)=-12\pi \rm{Im} \Pi^\text{EM}(-s+i\epsilon)$
is related to the QCD
expression for the  total cross-section of the $e^+e^-$-annihilation to a hadrons
process as $R(s)=\sigma_{tot}(e^+e^-\rightarrow {hadrons})/
\sigma(e^+e^-\rightarrow \mu^+\mu^-)$, where
$\sigma(e^+e^-\rightarrow \mu^+\mu^-)=4\pi \alpha_e^2/(3s)$ is
the theoretical normalization factor.

Following  Eq.(\ref{NS-SI}) one can decompose the  Adler function to the NS and SI
parts as
\begin{equation}
\label{DEMSI}
D^\text{EM}(L,a_s)=D^\text{NS}(L,a_s)+D^\text{SI}(L,a_s)\,.
\end{equation}
This function is the renormalization-group (RG) invariant quantity and
therefore it obeys the standard RG equation {\it without} anomalous dimension,
 namely
\begin{equation}
\label{RGINV}
\bigg(\frac{\partial}{\partial L} +\beta(a_s)\frac{\partial}{\partial a_s}
\bigg)D^\text{EM}(L,a_s)=0
\end{equation}
where the QCD $\beta$-function  is
\begin{equation}
\label{eq:beta}
\mu^2\frac{\partial a_s(\mu^2)}{\partial \mu^2}=
\beta(a_s)=-a_s^{2} \sum_{i\geq 0} \beta_i a_s^{i}\,.
\end{equation}
In the normalization conditions we used  $a_s=\alpha_s/(4\pi)$;
the first coefficient of this function reads
$\beta_0 = (11/3)C_\text{A} - (4/3) T_\text{R}n_f$ .
The solution of Eq.(\ref{RGINV})  reads:
\begin{equation}
 \label{eq:RGI}
D^\text{EM}\left(L;a_s(\mu^2)\right)=D^\text{EM}\left(0;a_s(Q^2)\right)\,.
\end{equation}
We shall consider here the NS contribution to Eq.(\ref{DEMSI}),
which we  write down \cite{MS04,Kataev:2014jba} as
\begin{equation}
\label{DNSdefinition}
D^{\rm NS}(a_s(Q^2))=1+\sum_{i\geq 1}d_l^\text{NS}a_s^{l}(Q^2)=1+d^\text{NS}_1 \cdot \sum_{l\geq 1}d_la_s^{l}
(Q^2)\,,
\end{equation}
where
 the coefficients $d^\text{NS}_l$ are considered   within the class of the
MS-like ultraviolet (UV) schemes.
The coefficients $d_l=d^\text{NS}_l/ d^\text{NS}_1$, the overall normalization factor
$d^\text{NS}_1=3{\rm C_F}$ that is more appropriate for the BLM like
optimization.
The $\beta$-expansion representation introduced in \cite{MS04}
 prescribes to decompose these  coefficients in the last equation
of Eq.(\ref{DNSdefinition}) in the following way (see \cite{Kataev:2010du} as
well):
\begin{subequations}
\label{eq:d_beta}
\begin{eqnarray}
\label{eq:d_1}
d_1&=&d_1[0]=1\, , \\
d_2&=&\! \beta_0\,d_2[1]
  + d_2[0]\, ,\label{eq:d_2}\\
  d_3
&=&\!
  \beta_0^2\,d_3[2]
  + \beta_1\,d_3[0,1]
  +       \beta_0 \,  d_3[1]
  + d_3[0]\, ,\label{eq:d_3} \\
  d_4
   &=&\! \beta_0^3\, d_4[3]
     + \beta_1\,\beta_0\,d_4[1,1]
     + \beta_2\, d_4[0,0,1]
     + \beta_0^2\,d_4[2]
     + \beta_1  d_4[0,1]
     + \beta_0\,d_4[1] \nonumber \\
   && \phantom{\beta_0^3\, d_4[3]+ \beta_1\,\beta_0\,d_4[1,1]+ \beta_0^2\,}~+d_4[0]\,,
       \label{eq:d_4} \\
   &\vdots& \nonumber \\
 d_{N}
   &=&\! \! \! \! \! \!~~\beta_0^{N-1}\! d_{N}[N\!-\!1]+ \cdots + d_N[0]\,,
\label{eq:d_n}
\end{eqnarray}
\end{subequations}
where $\beta_i$ are the coefficients of the $\beta$-function in Eq.(\ref{eq:beta});
the notation $i_0, i_1,\ldots$  for the arguments of $d_n[i_0,i_1,\ldots]$ denotes the powers of $\beta_0, \beta_1,\ldots$.
The decompositions in Eqs.(\ref{eq:d_beta}) should contain all information
about strong charge renormalization by means of using there all the possible
$\beta_i$-terms.
For the reader's convenience
we present the  explicit forms of the decomposition in
(\ref{eq:d_2},~\ref{eq:d_3}) in Appendix A
for the case  of  QCD supplemented with multiplets of  MSSM gluinos.

 As  follows from the studied  in \cite{Chetyrkin:1980sa,Chetyrkin:1980pr}
renormalization prescriptions, which will be presented  in  details in Sec.3,
the expression for the Adler $D$-function in   Eq.(\ref{DEMSI}) can also be written
in the following form \cite{Baikov:2012zm}:
\ba
 D^\text{EM}(L,a_s) =  12\pi^2\, \left(\gamma^\text{EM}_{\rm ph}(a_s)
-
 \beta( a_s) \frac{d }{d a_s}\Pi^\text{EM}(L,a_s)
\right),
\label{eq:Adler-master}
 \ea
where $\gamma^\text{EM}_{\rm ph}(a_s)$ is the QCD
anomalous dimension of the photon  vacuum polarization, which
is defined as
\begin{equation}
\label{gammaEM}
\gamma^\text{EM}_{\rm ph}(a_s)=
\mu^2\frac{d}{d \mu^2}\Pi^\text{EM}(L, a_s)\Big|_{L=0}~~~.
\end{equation}
In the discussions below we
shall clarify that the coefficients of the photon anomalous dimension
$\gamma^\text{EM}_{\rm ph}(a_s)$ should not be neglected in the process
of construction of the $\{\beta\}$-expansion formalism of \cite{MS04} and
of the careful construction of
the   PMC scale-setting  prescription for the $D$-function (for
the consideration of this topic at the  next-to-next-to-leading order (NNLO)
see  \cite{Kataev:2014jba} as well).

\section{The Adler function calculations in QCD }
\label{Sec:3}
\subsection{The renormalization relations for the Adler function in QCD}
\label{Sec:3.1}
In our further discussions we shall  use the MS-like scheme  renormalization
prescriptions   for the  Adler $D$-function, which were described
in detail in \cite{Chetyrkin:1980sa,Chetyrkin:1980pr}.
They were used in the process of evaluation of the 2nd order
 perturbative QCD
correction to the $D$-function and to its spectral density
$R(s)$ (in brief the result was   published  in \cite{Chetyrkin:1979bj}).
The same renormalization prescriptions were used for the calculations
of the 3rd and 4th order QCD corrections to $\Pi^\text{EM}$ and $D^\text{EM}$
 in \cite{Gorishnii:1990vf,Baikov:2012zm}.
Following these prescriptions one should consider first
the renormalization equation for the inverse photon propagator
in QCD
\begin{equation}
\label{prescription}
1+ a\Pi^\text{EM}(L,a_s)=Z_\text{ph}\bigg(1+a_{B}\Pi_{B}^\text{EM}(L,a_{sB})\bigg)~~~.
\end{equation}
Here $Z_\text{ph}$ is the photon renormalization constant,
$\Pi^\text{EM}(L,a_s)$ and $\Pi_{B}^\text{EM}(L,,a_{sB})$ are the
renormalized and  unrenormalized (bare) photon
polarization functions, respectively.
Due to the Ward identity,
the bare electromagnetic  coupling $a_{B}$ is related to the
renormalized one $a$ as $Z_\text{ph}a_{B}=a=\alpha_e/(4\pi)$,
 while  the  QCD bare  coupling $a_{sB}$ is related to the
renormalized one  as $Z_{a_s}^{-1} a_{sB}=a_s=\alpha_s/(4\pi)$.
In the class of MS-like schemes $Z_\text{ph}$
has the following form:
\begin{equation}
 \label{eq:Z}
Z_\text{ph}=1+a\cdot Z=1+ a \cdot \sum_{l=1}a_s^{l-1} \sum_{k=1}^{l} Z_{l,-k} \varepsilon^{-k}~,
\end{equation}
where  $\varepsilon=(4-d)/2$ is the parameter of the dimensional regularization,
$a_s$ -- renormalized strong coupling.
The QCD expression for the bare photon  polarization function
$\Pi_{B}^\text{EM}(L,a_{sB})$ in Eq.(\ref{prescription}) reads
\begin{equation}
\label{eq:PiB}
\Pi_{B}^\text{EM}\left(L,a_{sB}\right)=\sum_{j=1}^{\infty}a_{sB}^{j-1}\exp(Lj\varepsilon)
 \sum_{i=-j}^{\infty}
\Pi_{j,i}~\varepsilon^{i},
\end{equation}
where $\Pi_{j,i}$ are
 the elements of expansion of $\Pi^\text{EM}_B$ in the double series.
To get the renormalized expression for $\Pi^\text{EM}(L,a_s)$,
which determines eventually the renormalized Adler function,
one would use  $Z_\text{ph}a_{B}=a$ (the Ward identity) for electromagnetic coupling
and rewrite Eq.(\ref{prescription})
in terms of  $Z=(Z_\text{ph}-1)/a$:
\begin{equation}
\label{redifinition}
\Pi^\text{EM}(L,a_s)= Z  + \Pi_{B}^\text{EM}(L,a_{sB})\,.
\end{equation}
We take  into account
the known  relation between the bare coupling $a_{sB}$ and renormalized
one $a_s$ at the $O(a_s^3)$ level,
\begin{equation}
\label{eq:asB}
a_{sB}=a_s+a_s^2(-\beta_0/\varepsilon)+a_s^3(\beta_0^2/\varepsilon^2-\beta_1/\varepsilon)
+O(a_s^4),
\end{equation}
and then substitute the expression for $Z$ in Eq.(\ref{eq:Z}) and Eq.(\ref{eq:PiB}) together with
Eq.(\ref{eq:asB}) into the RHS of Eq.(\ref{redifinition})
 to obtain various relations between the pole parts $Z_{l,-k}$ and the elements $\Pi_{j,i}$.
  Using these relations in the definition of Eq.(\ref{D-function}) for $D^\text{EM}$
  leads \cite{Chetyrkin:1980sa,Chetyrkin:1980pr} to the following  expressions for the first
  perturbation coefficients \textbf{ $d_i^{NS}$}
\begin{subequations}
 \label{eq:d_n}
\begin{eqnarray}
d_1^\text{NS} &=&  - 2Z_{2,-1}, \label{eq:d-1}\\
d_2^\text{NS} &=&  - 3Z_{3,-1}+\beta_0\Pi_{2,0}.  \label{eq:d-2} \\
\text{Further application of}\!& &\! \text{
the same renormalization prescriptions gives} \nonumber \\
d_3^\text{NS} &=&  - 4Z_{4,-1}+2\beta_0\Pi_{3,0}+\beta_1\Pi_{2,0}+2\beta_0^2\Pi_{2,1}
\label{eq:d-3}.
\end{eqnarray}
 \end{subequations}
The latter was obtained in \cite{Gorishnii:1990vf}.
As we have already discussed after Eq.(\ref{eq:asB}),
the RHS of Eqs.(\ref{eq:d-1})-(\ref{eq:d-3}) can be rewritten also in terms
$\Pi_{j,i}$
instead of $Z_{l,-1}$, namely
\begin{subequations}
 \label{eq:dPi}
 \begin{eqnarray}
\label{eq:pi-d-1}
&&d_1^\text{NS} =   2\Pi_{2,-1},\\
&&d_2^\text{NS} =   3\Pi_{3,-1}- 2\beta_0\Pi_{2,0} \label{eq:pi-d-2}\,, \\ %
&&d_3^\text{NS}=   4\Pi_{4,-1} - 6\beta_0\Pi_{3,0} -\beta_1\Pi_{2,0} +6\beta_0^2\Pi_{2,1} \label{eq:pi-d-3}\,. \end{eqnarray}
\end{subequations}
In the discussions below we will prove that Eqs.(\ref{eq:d-1})--(\ref{eq:d-3}),
which contain the coefficients of the
photon anomalous dimension function of Eq.(\ref{gammaEM}), i.e. the
terms $-lZ_{l,-1}$, give the $\{\beta\}$-expanded structure
for $d_{l-1}^\text{NS}$ coefficients, which is  identical to the one formulated  in
\cite{MS04}  (see, e.g.,  Eqs.(\ref{eq:d_1})--(\ref{eq:d_3})
presented above).

\subsection{The  Adler function  in terms of photon anomalous dimension $\gamma^\text{EM}_{\rm ph}$}
\label{Sec: 2.3}
 Here we consider the results for $d^\text{NS}$ in Eqs.(\ref{eq:d-1} - \ref{eq:d-3})
 from the point of view of the general formula, Eq.(\ref{eq:Adler-master}).
We start with the expansion of the renormalized polarization function,
which follows from the definition (\ref{NS-SI})
\begin{equation}
\label{Piexp}
\Pi^\text{EM}(a_s)= \Pi^\text{EM}(L, a_s)|_{L=0}\equiv
\frac{d_R}{(4\pi)^2} \sum_{i\geq 1} a_s^{i-1}~\Pi^\text{R}_i,
\end{equation}
where we supply the coefficients of this expansion with superscript R to distinguish
them from the ones in expansion (\ref{eq:PiB}) of the bar $\Pi^\text{EM}_{B}$.
The corresponding anomalous dimension of the photon vacuum polarization
function, already defined in Eq.(\ref{gammaEM}),
\begin{equation}
\nonumber
\Ds \gamma^\text{EM}_{\rm ph}(a_s)=
\mu^2\frac{d}{d \mu^2}\Pi^\text{EM}(L, a_s)\Big|_{L=0}\,,
\end{equation}
can be derived in the MS-scheme from the first pole coefficient $Z_{-1}(a_s)$ of $Z$ in the expansion (\ref{eq:Z})
\begin{equation}
\label{ed:gamma-ph}
\gamma^\text{EM}_\text{ph}(a_s)= -\partial_{a_s}\left[a_s\sum_{l\geq 1} a_s^{l-1} Z_{l,-1}\right]=
\sum_{l\geq 1} a_s^{l-1}~\left(-l Z_{l,-1}\right)=
\frac{d_R}{(4\pi)^2} \sum_{j\geq 0}a_s^j~\gamma_j\,.
\end{equation}
Thus, the first terms in the RHS of Eqs.(\ref{eq:d-1}--\ref{eq:d-3}), namely
$- 2Z_{2,-1}, - 3Z_{3,-1} $ , $-4Z_{4,-1}$,
are  the perturbation theory  coefficients of $\gamma^\text{EM}_\text{ph}$
defined in the RHS of Eq.(\ref{ed:gamma-ph}).
Note that the sum $\sum_{l\geq 1} a_s^{l-1} Z_{l,-1}$ in the LHS of Eq.(\ref{ed:gamma-ph})
forms the coefficient $Z_{-1}(a_s)$.
The  analytical expressions for  $\gamma_i $ and
$\Pi^\text{R}_j$, were calculated in the \MSbar-scheme up to $i=4,~j=4$ in \cite{Baikov:2012zm}.
Both terms $\gamma^\text{EM}_{\rm ph}(a_s)$ and $\Pi^\text{EM}(a_s)$ in the RHS of
Eq.(\ref{eq:Adler-master}) contain the traces of $a_s$ renormalization
accumulated in the $\beta$-function (more precisely in its coefficients).
 In contrast  to the conclusion,
 presented in \cite{Wu:2013ei, Brodsky:2013vpa, Ma:2015dxa},
both these terms should be taken into account in the $\{\beta\}$-expansion.
This expansion  of $D^\text{EM}$ in order of $a_s^2$ has already been discussed
in Sec.3.3 of \cite{Kataev:2014jba},
where the values for $\Pi^\text{R}_2$ and $\gamma_3$
from \cite{Baikov:2012zm} were used
(there, the numeration
of $\Pi^\text{R}_i$ in index $i$ is shifted by 1 less and coupling
constant is $a_s=\alpha_s/\pi$).
There was already shown that the $\beta_0$-part of $\gamma_2= - 3Z_{3,-1}$
(these contributions  are underlined in the expression presented below)
together with the second term $\beta_0 \Pi^\text{R}_2$ enters into the
coefficient of $d_2^{NS}$ in  Eq.(\ref{eq:d-2}) as
\begin{subequations}
 \ba
\!\!d_2^\text{NS}= \underline{\frac{3}{4} \gamma_2} +  \frac{3}{4}~\beta_0\Pi^\text{R}_2 &=&\!\!
3{\rm C_F}\cdot \left[\underline{\left(\frac{\rm{C_A}}3-\frac{\rm{C_F}}2+
\frac{11}{12}\beta_0\right)}+\beta_0\left(\frac{55}{12}-4\zeta_3\right) \right]\equiv
\label{eq:gamma_2beta}\\
 3{\rm C_F}\cdot d_2&=&\!\! 3{\rm C_F}\cdot \left[\left(\frac{\rm{C_A}}3-\frac{\rm{C_F}}2\right)
+\beta_0\left(\frac{11}{2}-4\zeta_3\right) \right],\label{eq:d_2beta}
\ea
 \end{subequations}
and is consistent with  the standard BLM result
\cite{Brodsky:1982gc}, which is given by  Eq. (\ref{eq:d_2beta}).
Indeed,
the presentation of $D^\text{EM}$ as the decomposition (\ref{eq:Adler-master})
has a rather ``technical'' sense,
and for partial orders they can be presented in another form.
For instance,
in  the expressions in the RHSs of Eqs.(\ref{eq:dPi})
the anomalous dimension terms do not appear explicitly.
It happens that in contrast to the initial polarization
function $\Pi^\text{EM}(L, a_s)$  that
defines  in  Eq.(\ref{gammaEM}) the  anomalous dimension
$\gamma^\text{EM}_\text{ph}$,
the RGI quantity $D^\text{EM}(L,a_s)$ does not have its own
anomalous dimension.
  Therefore, there is no  physical reason to prefer
  only the second term of Eq.(\ref{eq:Adler-master})
(or, the second term in Eq.(\ref{eq:gamma_2beta})),
considering them as a unique origin of the $a_s$ renormalization,
as was done in, \eg, \cite{Brodsky:2013vpa,Ma:2015dxa}.

A similar expression for $d_3^\text{NS}$, which
follows from Eq.(\ref{eq:Adler-master}),  reads
 \ba
\!\!d_3^\text{NS}\equiv 3{\rm C_F}\cdot d_3&=&
\frac{3}{4}\left[ \gamma_3 + \beta_1 \Pi^\text{R}_2 + \beta_0 2 \Pi^\text{R}_3\right] =
\frac{3}{4} \left[ \gamma_3 + \beta_1 \Pi^\text{R}_2+
\beta_0\left(\beta_0 2\Pi^\text{R}_{3\beta} +2 \Pi^\text{R}_{30}\right) \right] . \label{eq:d_3beta}
\ea
The  contribution of $\Pi^\text{R}_2$ and
of   the decomposed  parts of the  term $\Pi^\text{R}_3=\beta_0\Pi^\text{R}_{3\beta} +\Pi^\text{R}_{30}$ in
Eq.(\ref{eq:d_3beta}) can be extracted directly from the
results  for $\Pi^\text{R}_i$ in \cite{Baikov:2012zm}\footnote{The analytical  expressions
for the QED contributions to $\Pi^\text{R}_2$,  $\Pi^\text{R}_3$ were
earlier presented in \cite{Gorishnii:1991hw}.}
and reads
\begin{subequations}
\label{eq:d_3betaPi}
\ba
\Pi^\text{R}_2= 2{\rm C_F} \left(\frac{55}{12}-4\zeta_3\right);~\Pi^\text{R}_{3\beta}=
  {\rm C_F} \left(\frac{3701}{54}-\frac{152}3\zeta_3\right);~\\
  \Pi^\text{R}_{30}= 2{\rm C_F} \left[{\rm C_F}\left(-\frac{143}9-\frac{148}3\zeta_3+80\zeta_5\right)+
  {\rm C_A}\left(\frac{146}{13}-8\zeta_3-\frac{40}{3}\zeta_5 \right)
   \right]\,.
\ea
\end{subequations}
Note that the   coefficient $\gamma_3$ of the photon anomalous
dimension function of Eq.(\ref{ed:gamma-ph})
has its own $\beta$-expansion, namely
\ba
\gamma_3&=&\!\! \beta_0^2\,\gamma_3[2]
  + \beta_1\,\gamma_3[0,1]
  + \beta_0\,\gamma_3[1]
  + \gamma_3[0]\,,   \label{eq:gamma_3beta}
\ea
which will be discussed in Sec.\ref{Sec: R'-operation}.
Therefore, the $\{\beta\}$-expanded
result for $d_3^\text{NS}$ in Eq.(\ref{eq:d_3beta})
(or $d_3$ in Eq.(\ref{eq:d_3}) )
is composed from the $\{\beta\}$-expansion
of both parts of Eq.(\ref{eq:Adler-master}).
To complete the consideration, we present here the
explicit expressions for the
elements of the $\{\beta\}$-expansion of  $\gamma_3$,
 defined in Eq.(\ref{eq:gamma_3beta}),
in the case when   QCD  is  supplemented with the MSSM multiplet of gluino:
\begin{subequations}
\label{eq:gamma}
\ba
  \gamma_3[2]&=&-{\rm C_F}\frac{77}{27}=\frac{4}3\cdot\left[d_3^\text{NS}[2]-C_\text{F}\left(\frac{3701}{36}-76\zeta_3\right)\right], \label{eq:gamma2}\\
  \gamma_3[1]&=&4{\rm C_F}\left[{\rm C_A}\left(-\frac{1249}{108}+\frac{104}3\zeta_3 \right)-
   {\rm C_F}\left(\frac{19}{9}+\frac{8}3\zeta_3 \right)\right] ,\\
 \gamma_3[0,1]&=&4C_\text{F}\left(\frac{23}{6} -4\zeta_3 \right), \\
  \gamma_3[0]&=&4C_\text{F}\left[ \left(\frac{523}{36}-
        72 \zeta_3\right){\rm C_A^2}
    +\frac{71}3 {\rm C_A C_F} - \frac{23}{2} {\rm C_F^2}\right]= \frac{4}3\cdot d_3^\text{NS}[0]\,.\label{eq:gamma0}
\ea
\end{subequations}
The RHSs of Eqs.(\ref{eq:gamma}) are derived from the result for
$\gamma_{ph}(n_f,n_{\tilde{g}})$, obtained in \cite{Chetyrkin:1996ez}
in the same way as it was done
 for the elements of the
$\{\beta\}$-expansion of the $d_3$ coefficient of the
Adler
function determined  in \cite{MS04,Kataev:2010du,Kataev:2014jba}
(these elements and the required $\beta$-function coefficients are presented
in the  Appendix).

The value in the first equation (\ref{eq:gamma2}) can be easily checked from
the explicit result in \cite{Baikov:2012zm}
(see Eq.(3.13) there).
The last equation in (\ref{eq:gamma0}) has a general reason:
only $\gamma_n[0]$-terms  determine the expressions  for $d_n^\text{NS}[0]$.
 This feature can be clearly seen from the expression  of Eq.(\ref{eq:Adler-master}).
It also
clarifies that the $\beta$-dependent contributions to the elements defined in
(\ref{eq:d_beta}) are formed  by the corresponding  contributions from both
$\gamma^\text{EM}_\text{ph}$ and $\Pi^\text{EM}$.
For example, substituting
Eqs.(\ref{eq:gamma}) into  Eq.(\ref{eq:d_3beta}) one can obtain
the following  structure of the $d_3^\text{NS}$ -term:
 \ba
\!\!d_3^\text{NS}&=&
\frac{3}{4}\left[ \gamma_3[0] + \beta_0\left(\gamma_3[1]+2\Pi^\text{R}_{30}\right)+\beta_1
\left(\gamma_3[0,1]+ \Pi^\text{R}_2\right) + \beta_0^2 \left(\gamma_3[2] +2\Pi^\text{R}_{3\beta}\right)\right]\,,
\label{eq:d3NS}
\ea
where
\ba
d_3^\text{NS}[2]=\frac{3}{4}\left(\gamma_3[2]+2\Pi^\text{R}_{3\beta}\right),
~d_3^\text{NS}[0,1]=\frac{3}{4}\left(\gamma_3[0,1]+\Pi^\text{R}_2\right),
~d_3^\text{NS}[1]=\frac{3}{4}\left(\gamma_3[1]+2\Pi^\text{R}_{3,0}\right)\,. \nonumber
\ea
 In   Sec.\ref{Sec: R'-operation}, we shall discuss the
origin  of this decomposition using the language of the
 incomplete BPHZ  $R$-operation, i.e., the  $R'$-operation.

\section{The structure of the $R'$-operation and the $\{\beta \}$-expansion}
 \label{Sec: R'-operation}
Our goal here is  to trace the 
 form of the expansion (\ref{eq:d_3}) in regard to
the term $\gamma_3$ of anomalous dimension $\gamma^\text{EM}_\text{ph}$ that, in turn, enters into $d_3$ by means of presentation
(\ref{eq:Adler-master}, \ref{eq:d3NS}).
 The method we use is more general and relates the structure of the $\{\beta \}$-expansion
 to the structure of various contractions of the $R'$-operation for \textit{any} perturbative order.
 To be more precise these $R'$-operation contractions of the subgraphs form the
 corresponding contributions to the $\beta_i$ coefficients of the $\beta$--function.

 We start with
the  definition of the renormalization constant $Z$ in the MS-scheme \cite{Vladimirov:1977ak},
 \be
 Z_{}=1 - \hat{K}R'(G),
 \ee
where $G$ denotes the set of the corresponding diagrams
(or a single one in the case of partial contribution) and
$\hat{K}$ separates out poles in $\varepsilon$.
One can rewrite expression (\ref{gammaEM}) for the anomalous dimension $\gamma^\text{EM}_\text{ph}$: 
\ba
\gamma^\text{EM}_\text{ph}=-\partial_{a_s}\left[a_s Z^{(1)}_{}\right] =
\partial_{a_s}\left[a_s \hat{K}_1 R'(G) \right]= 4\hat{K}_1 R' (G)\Big|_\text{4-loops}\, ,
\label{Z}
\ea
 where the last equality is written for the three-order contribution $a_s^3 \gamma_3$.
Here $R'$ -- the incomplete BPHZ $R$-operation, $R=R'-\hat{K} R'$  \cite{BSHbook},
\ie, the $R'$-operation subtracts all the subdivergences of internal
non-intersecting subgraphs but does not subtract an overall divergence $\hat{K} R'$ of a diagram,
see, \eg,
\cite{Chetyrkin:1980pr,Tarasov:2013zv};
$Z^{(1)}= -\hat{K}_1 R'(G)$
where $\hat{K}_1$ picks out the coefficient at the first pole.
We will not present here the definition of the well-known $R'$-operation,
instead of that we will refer an interested reader to the detailed article \cite{Tarasov:2013zv}
containing a number of appropriate examples
 and  notations,  which  are used below.

One can classify the origins of different contributions to $\gamma_3$
in the expansion 
\ba  \label{eq:gamma_beta}
\gamma_3= \beta_0^2\,\gamma_3[2]
  + \beta_1\,\gamma_3[0,1]
  + \beta_0\,\gamma_3[1]
  + \gamma_3[0]\, ,\nonumber
  \ea
and  relate
this $\beta$-expansion
to the elements of the structure of the $(R'-\1)$ operation in the last
equality in the RHS of Eq.(\ref{Z}).
In this notation the contribution to the $\beta$-function can be obtained as
\be \label{eq:defbeta}
\beta(a_s)= a_s^2\partial_{a_s}Z^{(1)}_g=a_s^2\partial_{a_s}\left[-\hat{K}_1 R' (G^g) \right]=-a_s j \left[\hat{K}_1 R'(G^g) \right],
\ee
where $Z_g$ is a contribution to the charge renormalization constant and
$j$  is  the  power
in $a_s$,  which enter   the subgraph $G^g$.
Further, we shall classify different contributions to $\gamma_3$ following
different kinds of contractions of subgraphs of the corresponding 4-loops diagrams $G_4$.
Every contraction of the subgraph $G^{(l)}$ in $\hat{K}_1 R' (G_4)$ is related
to either   the
renormalization of the charge $g$
($g^2/(4\pi)=\alpha_s = 4\pi a_s$)  where  $G^{g}$
 is the
subgraph that finally contributes to
$\beta(a_s)$ through $Z^{(1)}_g$ in Eq.(\ref{eq:defbeta}),
or   the renormalization of electromagnetic vertex $J^\text{EM}_{\mu}$
and  of its legs,  with the related     $G^{J}$ subgraphs.
The later renormalizations  cancels
in the sum due to the Ward identities (WI) \cite{Chetyrkin:1980sa}.

Let us  clarify this  more formal  proof considering
 all possible contraction of the subgraphs.

(i)
Diagrams $G_4$ may admit only 2 contractions of the 1-loop intrinsic subgraphs $G_1^{(l)}$
under the $R'$-operation.
Let us choose those subgraphs, $G^{g}$, that renormalize the
intrinsic charges $g$ thus form the renormalization constant $Z_g$,
see the upper part in Fig.\ref{fig:rest1,1}.
 The residual of any diagrams under these contractions, $\Gamma^\text{rest(1,1)}$,
 reduces to sum of three diagrams shown in the lower part in Fig.\ref{fig:rest1,1}.
 \begin{figure}[th]
 \includegraphics[width=0.28\textwidth]{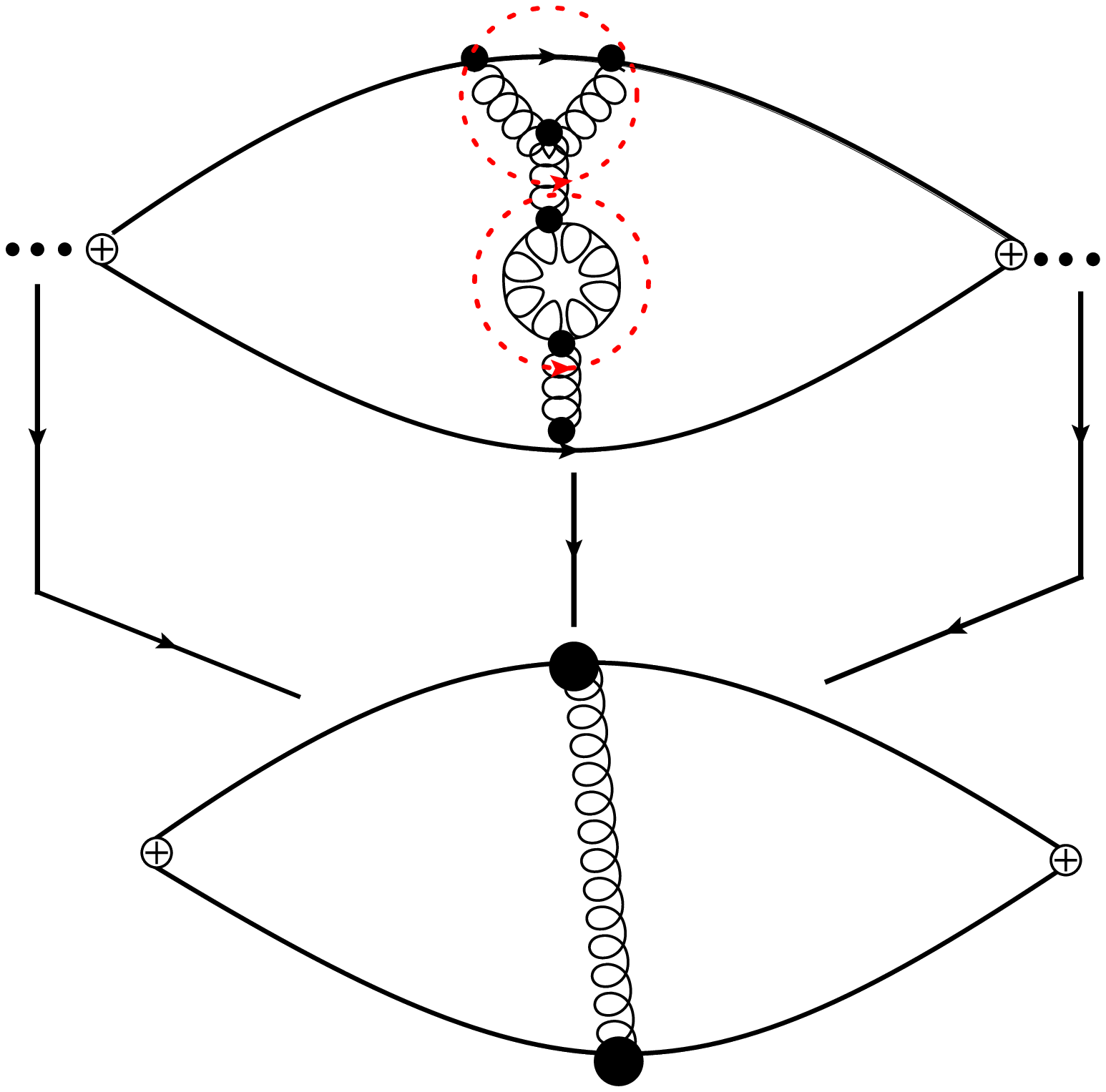}~
 \includegraphics[width=0.28\textwidth]{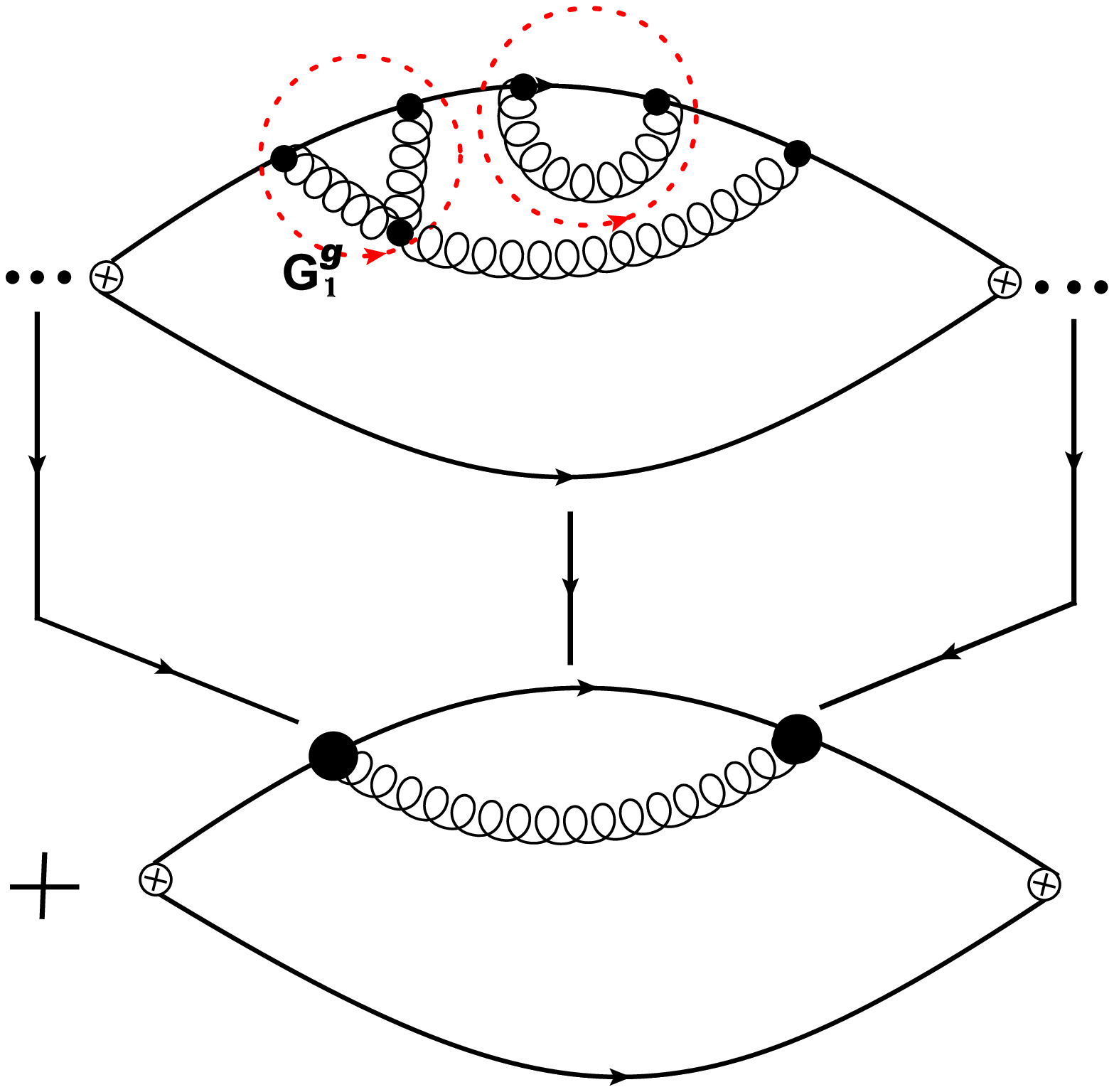}~
 \includegraphics[width=0.28\textwidth]{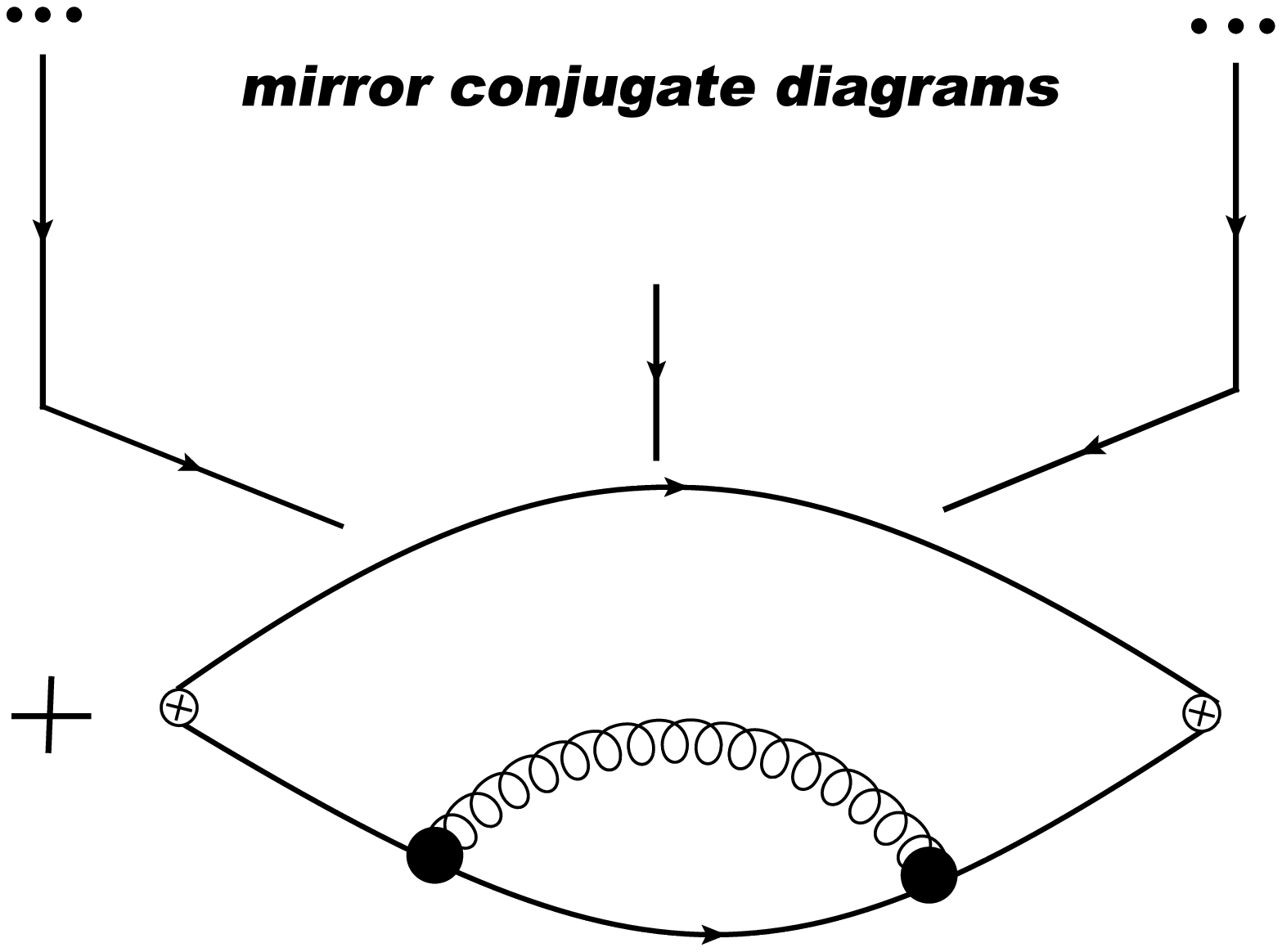}
  \caption{\label{fig:rest1,1} \footnotesize{
Upper row: the samples  of 1-loop contractions, the dashed circle
(in red) around
graph $G$  denotes the operation
 $-\hat{K}R'(G)$; here $\oplus$ is the electromagnetic vertex
$J^\text{EM}_{\mu}$. Lower row: the diagrams, which determine the reminder
$\Gamma^\text{rest(1,1)}$ after two of 1-loop  $G_1$
contractions of the  4-loop diagrams $G_4$.}
}
  \end{figure}

This residual $\Gamma^\text{rest(1,1)}= G_4\setminus G_1^{g(l)}\setminus G_1^{g(m)}$,
resulting\footnote{Sign $" \setminus "$ in an expression $" A\setminus B~"$ means
the subtractions of the subgraph $B$ from the graph $A$}
after two  one-loop contractions of $G_4$, is unique
 for any kind of contractions.
 The sum $ \sum_{l}(\hat{K}G_1^{g(l)}) \sim \beta_0$ (here $R'G_1=G_1$);
 therefore, the contribution to $\gamma_3$ after 2 contractions is
 \ba \label{eq:rest-1-1}
  \left[ \Ds
   \hat{K}_1\sum_{m,l}\Gamma^\text{rest(1,1)}(\varepsilon)\cdot\left(-\hat{K}G_1^{(l)}\right)
 \left(-\hat{K}G_1^{(m)}\right) \sim
  \Gamma^\text{rest(1,1)}_{\varepsilon}\cdot \beta_0^2 \right]
 \stackrel{\rm contr.(1,1)}{\longrightarrow} \gamma_3[2]\cdot \beta_0^2,
 \ea
 where
 $\Gamma^\text{rest(1,1)}_{\varepsilon}=\left(\Gamma^\text{rest(1,1)}(\varepsilon)\right)'|_{\varepsilon=0}$.
 After the sum over all contractions $(i,j)$ one obtains the factor $\beta_0^2 $.
 The contributions of those one-loop  contracted subgraphs that include   $J^\text{EM}$ vertices
  are cancelled by the contracted self-energy parts of fermions due to WI.
  The mixed cases including the contractions of the $G_1^{J}$ and $G^{g}$ subgraphs should
  result  in zero by the same reason.
 Finally, the term $\Gamma^\text{rest(1,1)}_{\varepsilon}$ contributes to the corresponding
 $\gamma_3[2]$ element of the $\beta$-expansion.

(ii)
The different contractions of the 2-loop subgraphs $G_2^{g(k)}$ that renormalize  charges,
lead to the $\beta_1$-part of the expansion at the single residual $\Gamma^\text{rest(2)}= G_4\setminus G_2^{g(k)}$
of the diagram
\ba
 \left[
  \hat{K}_1\sum_{k}\Gamma^\text{rest(2)}(\varepsilon)\left(-\hat{K}R'(G_2^{(k)})\right) \sim
 \Gamma^\text{rest(2)}_{0}\beta_1 + \Gamma^\text{rest(2)}_{\varepsilon} \beta_0^2 \right]
 \stackrel{\rm contr.(2)}{\longrightarrow} \gamma_3[0,1] \cdot \beta_1+\!\!\ldots,
\ea
  where $\Gamma^\text{rest(2)}_{0}=\Gamma^\text{rest(1,1)}(0)$ is the part of
  $\Gamma^\text{rest(2)}$ finite in $\varepsilon$.
 These terms contribute to the element $\gamma_3[0,1]$.
 Besides, the high pole ($1/\varepsilon^2$) of these contractions contribute to the term
  $\sim \Gamma^\text{rest(2)}_{\varepsilon}\cdot \beta_0^2 $, i.e., again to
  $\gamma_3[2]~\beta_0^2$.
  The contributions from the contracted $G_2^{J}$ subgraphs are factorized with the single residual given
  by the diagrams in the lower row of Fig.\ref{fig:rest1,1}.
 Due to the WI the sum is
zero,  \ie,  $\sum_{l}\left(-\hat{K}R'(G_2^{J(l)})\right) =0$.

(iii) All the contractions of the 3-loop subgraphs $G_3^{(k)}$ of $G_4$ become the
$G_3^{J(k)}$ subgraphs.
 One can verify
 that  in this
case the single residual $\Gamma^\text{rest(3)}(\varepsilon)$ is
given by the contribution of the simple quark loop $\Pi_1$,
while the sum $\sum_{m}\left(-\hat{K}R'(G_3^{J(m)})\right) =0$ again due to the WI.

(vi)
The contractions of the 1-loop subgraphs $G_1^{g(k)}$ that form
 $Z_g$  of a certain charge $g$ contribute to the $\beta_0$-term and form in part the element $\gamma_3[1]$,  namely
 \ba \label{eq:rest-1}
 \left[
  \hat{K}_1~\sum_{k}\Gamma^\text{rest(1k)}(\varepsilon)\cdot\left(-\hat{K}R'(G_1^{g(k)})\right) \sim
 \Gamma^\text{rest(1)}_{0}\cdot \beta_0 + \ldots \right] \stackrel{\rm contr.(1)}{\longrightarrow}
 \gamma_3[1]\cdot \beta_0 + \ldots.
\ea
Here $\Gamma^{\text{rest(1)}k}=G_4\setminus G_1^{g(k)}$ are the same for a set of subgraphs,
the common coefficient $\Gamma^\text{rest(1)}_{0}$ is a sum of the
$\Gamma^{\text{rest(1)k}}$  terms.
Of course these kinds of contributions can produce the $\beta_0^2$-term also,
if $\Gamma^\text{rest(1)}_{0}$ contains $\beta_0$.
The latter possibility is  indicated in (\ref{eq:rest-1}) by  dots.

The previous analysis is enough to fix the structure of the $\gamma_3$ coefficient,  but not
the values of $\gamma_3[\ldots]$ elements.
 We realize that
the results in the items (i) - (vi)
are not the single source of the $\beta$-terms in the expansion,
the term $\hat{K}_1(G_4)$
 formed from the subgraph  $G_4$ without any contractions
 contributes too.
 Therefore
\ba 
 \left[
  \hat{K}_1~\left( G_4\right) \sim
  \beta_0^2\cdot (\ldots) +\beta_1\cdot (\ldots) +\beta_0\cdot (\ldots)+ C\right]
  \stackrel{\rm contr.(0)}{\longrightarrow} \gamma_3
\ea
may contain any of the aforementioned elements of the $\beta$-structure
and, in addition, the   independent of $\beta_0$ and $\beta_1$
contribution,  denoted by $C$.
Finally, we can conclude that the structure of the $\{\beta \}$-expansion
 of $\gamma_3$
appears
as a natural result of different contractions $\left(-\hat{K}R'\left(G_{1,2,\ldots}^{g}\right)\right)$
of the subgraphs $G_{1,2,\ldots}^{g}$ under the
 action of
$R'$-operation.

\section{Generalized $\{ \beta,\gamma \} $--expansion}
\label{Sec:beta-gamma}
Another kind of the $\{ \beta \}$-expansion appears for  RG covariant (RGC) quantities,
those that have their own
 anomalous dimension
For these RG covariant  dimensionless one-scales quantities,
say Green function $S\left(Q^2/\mu^2;a_s(\mu^2)\right)$,
the well-known RG transform is
\begin{subequations}
 \label{eq:RGC}
\ba
S\left(Q^2/\mu^2; a_s(\mu^2)\right)&=&\hat{z}\left(t,a_s(\mu^2)\right) \cdot S\left(Q^2/\mu^2~\frac{1}{t};
~\bar{a}_s(t,a_s(\mu^2))\right), \label{eq:RGC-a}\\
\text{where}~ \hat{z}\left(t,a_s(\mu^2)\right)&=&\exp\left\{ \int_{1}^{t}\gamma_S(\bar{a}_s(\tau,a_s(\mu^2)) \frac{d\tau }{\tau}
\right\},~ \bar{a}_s(1,a_s(\mu^2))=a_s(\mu^2),
\label{eq:RGfactor}
\ea
 \end{subequations}
for the simplest case without
mixing\footnote{For the case of mixing the $\hat{z}$ and $\gamma_S$ are matrixes},
where the scale extension factor $t=\mu'^2/\mu^2$ and the
 anomalous dimensions
of $S$ is
$\gamma_S(a_s)= \sum_{j=1}a_s^j ~\gamma_j$.
Given $\gamma_S(a_s)=0$ one returns to the condition (\ref{eq:RGI}) for RGI quantities in Section \ref{Sec:2}.
Let us consider  the perturbative expansion of  $S\left(1; \bar{a}_s(Q^2)\right)$
at the value $\mu^2=Q^2$,  $s_0=1$,
\be \label{eq:S-series}
S\left(1; \bar{a}_s\right)= 1 + \sum_{n=1} \bar{a}_s^n s_n\,,
\ee
and let analyse the structure of the expansion coefficients $s_n$ in comparison with the RGI
case in Eq.(\ref{eq:d_beta}).
The decomposition of $s_n$ should include along with elements of the $\{ \beta \}$-expansion also
the elements with  $\gamma_j$ that we will separate from the first ones.
The difference with the standard $\{\beta \}$-expansion reveals itself starting with $s_1$,
which includes now the contribution with $\gamma_0$
and are marked  in (\ref{eq:s_beta}) by the bold font, while the standard $\beta$-terms --
by the italic one, namely
\begin{subequations}
\label{eq:s_beta}
\begin{eqnarray}
\label{eq:s_1}
\!\!\!\!s_1&=&s_1[0] + \gamma_0 \bm{s_1[0|1]}, \\
\!\!\!\!s_2&=& s_2[0] + \beta_0\,{\it s_2[1|0]}+
 \nonumber \\
  &&\phantom{s_1[0]} \gamma_1 \bm{s_2[0|0,1]} + \gamma_0^2 \frac{1}{2}\bm{s_2[0|2]} - \gamma_0 \beta_0 \frac{1}{2}\bm{s_2[1|1]}
  ,\label{eq:s_2}\\
   s_3
&=&\! s_3[0]+
  \beta_0^2\,{\it s_3[2|0]}
  + \beta_1\,{\it s_3[0,1|0]}
  +       \beta_0 \,  {\it s_3[1|0]}+
  \nonumber \\
 && \phantom{s_1[0]}\gamma_2 \bm{s_3[0|0,0,1]}-\frac{\gamma_0}2 \beta_1 \bm{s_3[0,1|1]} -\gamma_1 \beta_0 \bm{s_3[1|0,1]}\, -\gamma_1 \gamma_0 \bm{s_3[0|1,1]} + \nonumber\\
 && \phantom{s_1[0]} \frac{\gamma_0}{3} \beta_0^2 \bm{s_3[2|1]} - \frac{\gamma_0^2}{2} \beta_0 \bm{s_3[1|2]}+\frac{\gamma_0^3}{3}\bm{s_3[0|3]}   ,\label{eq:s_3} \\
 &\vdots&\,. \nonumber 
\end{eqnarray}
 \end{subequations}
The first two elements in the expansion of $s_2$ repeat the $\{ \beta \}$-expansion, while the other new admissible
 elements are proportional to $\gamma_{0}$ or $\gamma_{1}$.
 The coefficients ($1/2$) and sign are chosen in (\ref{eq:s_2})  in accordance with the
 perturbation expansion of the factor $\hat{z}(t, a_s)$ in Eq.(\ref{eq:z-expan}),
\begin{subequations}
  \ba
  \label{eq:z-expan}
   \hat{z}\left(t,a_s\right)&=&1+a_s \gamma_0 l + a_s^2\left(\gamma_1 + \gamma_0^2\frac{1}{2} l -\gamma_0\beta_0 \frac{1}{2}l
   \right) l+\\
  && a_s^3 \left[\gamma_2 - \left(\frac{\gamma_0}{2}\beta_1+\gamma_1(\beta_0 -\gamma_0) \right)l+ \frac{\gamma_0}3
  \left(\beta_0 -\frac{\gamma_0}{2} \right)(\beta_0 -\gamma_0)l^2 \right] l \label{eq:z-expan-3}\\
&&+ O(a_s^4);~~l=\ln(t)\,. \nonumber
\ea
 \end{subequations}
 The similar notations for the decomposition of $s_3$ in Eq.(\ref{eq:s_3}) follow to the expansion in (\ref{eq:z-expan-3}).
 All the admissible  elements of the RG generators ($\beta, \gamma_S$) for quantity  $S$
 should be taken into account in the decomposition of the coefficients $s_n$
 (some of these elements may be equal to zero) as the ``traces'' of the RG factor $ \hat{z}$.
 The notation we use in Eq.(\ref{eq:s_beta}) follows the rule:
 the first series $\{\beta \}$ before the separator in the square brackets of $s_n[\{\beta \}|\{\gamma \}]$
 means the number of powers of $\beta_i$
 in the $i$-position in full correspondence with the notation in Eq.(\ref{eq:d_beta});
 see, \eg, $\beta_0\,{\it s_2[1|0]}$ in (\ref{eq:s_2}), or  $\beta_1\,{\it s_3[0,1|0]}$ in (\ref{eq:s_3}).
 The second series $\{\gamma \}$ stays after this separator, it counts the powers of $\gamma_j$
 that accompanying this contribution, see, \eg, $\gamma_0\,\bm{s_1[0|1]}$.
 These new terms are presented in the second lines of (\ref{eq:s_2}), (\ref{eq:s_3}) and
 are constructed just in the same manner as the ones for $\beta_i$.
 So we separate here the $\{ \beta \}$-expansion from the $\{ \gamma \}$ one.

 The question is how we can use  this new detailed decomposition of the series coefficients?
For the purpose of the series optimization one can collect the terms with anomalous dimensions
$\gamma_i$ in decompositions in Eqs.(\ref{eq:s_beta}) and transfer them into the common $\hat{z}$ factor starting with,
\eg,
the \textit{first order} term $ \gamma_0 \bm{s_1[0|1]}$ in (\ref{eq:s_1}).
This fixes the logarithmic shift $\Delta$ \cite{Kataev:2014jba}, $\Delta=\bm{s_1[0|1]}=\ln(Q^2/\mu^2)$,
to the new normalization scale $Q^2$.
 The term $\beta_0 \Delta$ corresponded this new scale should be removed from the first line in Eq. (\ref{eq:s_2})
in the \textit{second order} and so on following to BLM procedure.
 We shall elaborate this kind of series optimization and apply them to certain  quantities somewhere else.
The main goal of this demonstration is to reveal the difference between the RGI and RGC quantities.
In the first case, one has the $\{ \beta \}$-expansion only,
while for the last case, one should distinguish the $\beta$ and $\gamma$--terms in the decomposition of
the expansion coefficients to form the common factor $\hat{z}$.

It should be mentioned that an optimization approach based on the
$\{ \beta,\gamma \} $-expansion should replace the PMC-type optimization approach for the concrete
case of the optimization of the total decay width of the $H^0\rightarrow b\overline{b}$ decay process,
considered in \cite{Wang:2013bla}.
The importance of this consideration was emphasized in \cite{Petrov:2015jea}.

\section{Conclusion}
We discussed  the $\{ \beta \}$-expansion for renormalization group (RG) invariant
quantities.
The origins of this expansion are traced  to the structure of the result of the incomplete
BPHZ $R$-operation -- the $R'$-operation,
 considering the role of various contractions there.
 All the coupling renormalizations following from these  contractions should be
  taken into account for the $\{ \beta \}$-expansion.
 We illustrate our  theoretical  conclusion  in the
$O(a_s^3)$ order
by means of analysis of the calculation scheme,  when  the nonsinglet Adler
  function $D^\text{NS}$ is expressed through the photon anomalous
dimension $\gamma_{Ph}^{EM}(a_s)$.
  Note here that in \cite{Cvetic:2016rot} another  special
 QCD  approach was used to determine
  the coefficients of the $\{ \beta \}$-expansion in Eqs.(\ref{eq:d_3},\ref{eq:d_4}).
  The detailed comparison of the results of its applications
with the results obtained here is on the agenda.

We proposed a generalization of the $\{ \beta \}$-expansion for the renormalization group
covariant (RGC) quantities -- the $\{ \beta,\gamma \}$-expansion.
This expansion can be the basis of a new optimization procedure for the RGC quantities.

\acknowledgments
We would like to thank S.~J. Brodsky and M.~Y. Kalmykov
for the fruitful discussion.

The work of A.K. was supported by the Russian Science Foundation
grant  No. 14-22-00161.
The work of M.S. was  supported in part by the
BelRFFR--JINR, grant F16D-004 and by   the
Russian Foundation for Basic Research, Grant No.\ 14-01-00647.
 \appendix
  \section{Explicit formulas for the elements of  $D$\\ and $\beta$-function coefficients}%
 \renewcommand{\theequation}{\thesection.\arabic{equation}}
\label{App:A}   \setcounter{equation}{0}
For the Adler function $D^\text{NS}$ obtained within QCD with light
gluinos $n_{\tilde{g}}$  the  elements
of the $\{ \beta \}$-expansion read
\begin{subequations}
\label{eq:d1-4}
 \begin{eqnarray}
d_1^\text{NS}&=&3{\rm C_F};~d_1=~1; \label{D-11}\\
d_2[1]&=&\frac{11}2-4\zeta_3;~~~~~
d_2[0]=\frac{\rm C_A}3-\frac{\rm C_F}2; \label{D-21} \\
d_3[2]&=&\frac{302}9-\frac{76}3\zeta_3;~d_3[0,1]=\frac{101}{12}-8\zeta_3;\label{D-32}\\
d_3[1]&=&
    {\rm C_A}\left(-\frac{3}4 + \frac{80}3\zeta_3 -\frac{40}3\zeta_5\right) -
    {\rm C_F}\left(18 + 52\zeta_3 - 80\zeta_5\right) \label{D-31}; \\
d_3[0]&=& \left(\frac{523}{36}-
        72 \zeta_3\right){\rm C_A^2}
    +\frac{71}3 {\rm C_A C_F} - \frac{23}{2} {\rm C_F^2}~.  \label{D-30}
\end{eqnarray}
\end{subequations}
The required $\beta$-function coefficients within the same scheme with the light
gluinos $n_{\tilde{g}}$ \cite{Clavelli:1996pz} and the number $n_f$ of the quark flavors calculated in the \MSbar scheme are
\begin{subequations}
 \label{eq:beta0-3}
\begin{eqnarray}
 \label{eq:beta-b0}
  \beta_0\left(n_f, n_{\tilde{g}}\right)
   &=& \frac{11}{3} C_A - \frac{4}{3}\left( T_R n_f + \frac{n_{\tilde{g}} C_A}{2}\right)\,;\\
 \label{eq:beta-b1}
  \beta_1\left(n_f, n_{\tilde{g}}\right)
     &=& \frac{34}{3}C_A^2
       - \frac{20}{3}C_A \left( T_R n_f + \frac{n_{\tilde{g}} C_A}{2}\right)
      -4\left( T_R n_f C_\text{F}+ \frac{n_{\tilde{g}} C_A}{2} C_A\right);\\
 \label{eq:beta-b2}
  \beta_2\left(n_f, n_{\tilde{g}}\right)
     &=& \frac{2857}{54}C_A^3
       - n_f T_R \left( \frac{1415}{27}C_A^2 +\frac{205}{9}C_A C_F -2 C_F^2 \right)
       + (n_f T_R)^2 \left( \frac{44}{9}C_F +\frac{158}{27}C_A \right) - \nonumber \\
       && \frac{988}{27}n_{\tilde{g}} C_A (C_A^2) +
       n_{\tilde{g}} C_A n_f T_R \left( \frac{22}{9}C_A C_F +\frac{224}{27}C_A^2 \right)
       +(n_{\tilde{g}} C_A)^2 \frac{145}{54} C_A \, .
\end{eqnarray}
 \end{subequations}
 The N$^3$LO coefficient $\beta_3\left(n_f, n_{\tilde{g}}\right)$ has been obtained recently in the papers \cite{Zoller:2016sgq,Bednyakov:2016uia}.

 For the sake of readers we present here also the $\gamma_{0,1,2}$ values from \cite{Baikov:2012zm},
rescaling them to our normalization of the coupling constant
$a_s=\alpha_s/(4\pi)$,
\ba
&&\gamma_0=\frac{4}3,~ \gamma_1=4{\rm C_F},~ \gamma_2=4{\rm C_F} \left(\frac{\rm C_A}{3}-\frac{\rm C_F}{2}+\frac{11}{12}\beta_0 \right).
\ea

\end{document}